\documentclass{article}
\date{ }

\title{THE MERCHANDISING  MATHEMATICIAN MODEL. STOCHASTIC DEMAND AND SUPPLY.}

\begin{document}
\maketitle
\begin{center}
By Edward W. Piotrowski and Jan S\l adkowski
\end{center}
\def\Z{{\bf Z\!\!Z}}
\begin{abstract}
A simple model of a buying-selling cycle is proposed. The model
comprises two moves: a rational buying and a random selling.  The
notion of a profit intensity is introduced. Supply and demand
curves and geometrical interpretation are discussed in this
context.
\end{abstract}


\newtheorem{theorem}{THEOREM}
\newtheorem{example}{EXAMPLE}
\section{INTRODUCTION}

The very aim of any conscious and rational economic activity is
optimization of the profit in given economic conditions and,
usually, during definite intervals. The interval is chosen so that
it contains a certain characteristic economic cycle (e.g. one
year, a season, an insurance period or a contract date). Of
course, often it is possible and reasonable to make prognosis for
a distant future of an undertaking by extrapolation from the
already known facts. The quantitative description of an
undertaking is extremely difficult when the time of duration of
the intervals in question is itself a random variable (denoted by
$\tau$ in the following). The profit gained during the specific
period, described as a function of $\tau$, becomes also a random
variable and as that does not measure the quality of the
undertaking. To investigate activities that have different periods
of duration we define, following the queuing theory Billingsley
(1979), the {\bf profit intensity} as a measure of this economic
category. An acceptable definition of the profit must provide us
with an additive function. It seems that the proposed interval
interest rate notion leads to consistent results.

\section{THE PROFIT INTENSITY}

Let $t$, $\upsilon _{t}$ and $\upsilon _{t+\tau}$ denote the
beginning of an interval of the duration $\tau $, the value of the
undertaking (asset) at the beginning  and at the end of the
interval, respectively. We  define the {\bf logarithmic rate of
return} $r_{t,t+\tau}$ as $$r_{t,t+\tau}\equiv \ln \left(
\frac{\upsilon _{t+\tau}}{\upsilon _{t}}\right). \leqno(1)$$ Let
the expectation value of the random variable $\xi$ in one cycle
(buying-selling or vice versa)  be denoted by $E\left( \xi
\right)$. If $E\left( r_{t,t+\tau} \right)$ and $E\left( \tau
\right)$ are finite then we define the {\bf profit intensity} for
one cycle $\rho _{t}$ Piotrowski(1999) as
$$\rho _{t} \equiv \frac{E\left(
r_{t,t+\tau} \right)}{E\left( \tau \right)}. \leqno(2)$$ This
definition is an immediate consequence of the Wald identity
Resnick(1998):
$$ E\left( S_{\tau'}\right)= E\left( X_{1}\right)E\left( \tau'\right)\ , \leqno(3)$$
where $S_{\tau'}\equiv X_{1}+\ldots +X_{\tau'}$ is the sum of $\tau'$
 equally
distributed random variables $X_{k},\ k=1,\ldots , \tau'$ and
$\tau'$ is the stopping time, Billingsley (1979) and Resnick
(1998). It is obvious that the profit intensity we have defined in
the Equation (2) is just the $E\left( X_{1}\right)$ from the Wald
identity, Equation (3). The expected profit is the left hand side
of the Wald identity. If we are interested in the profit expected
in a time unit we have, according to Wald, divide the expected
profit by the expectation value of the stopping time, so we get
the Equation (2). We can also calculate the variance of the profit
intensity by using the proposition 10.14.4 from the reference
Resnick (1998):
$$E\left( \Bigl(S_{\tau'} -\tau' E\left( X_{1}\right)\Bigr) ^{2}\right) =E\left(
\tau'\right) Var \left( X_{1}\right) . \leqno(4) $$ Of course, our
definition of the profit intensity is applicable also in more
general cases when the random variables $X_{i}$ are correlated or
have different distributions.

The profit expected after an arbitrary time interval, say $\left[
0,T\right]$ is given by
$$\rho _{0,T} \equiv \int _{0}^{T} \rho _{t}dt \ .  \leqno(5)$$

The proposed definition of the profit intensity is a convenient
starting point for the consideration of the proposed below model.
Relations to the commonly used measures of  profits (returns) can
be easily obtained by simple algebraic manipulations.

\section{THE  MERCHANDISING MATHEMATICIAN MODEL}

Let us consider the simplest possible market event of exchanging
two goods which we would call the asset and the money and denote
by $\Theta $ and $\$ $, respectively. The proposed model comprises
two moves. First move  consist in a rational buying of the asset
$\Theta $ (exchanging $\$ $ for $\Theta )$. The meaning of the
adjective rational will be explained below. The second move
consist in a random (immediate) selling of the purchased amount of the asset $
\Theta $ (exchanging $\Theta $ for $ \$ $). Note that the order of
these transactions can be reversed and, in fact, is conventional.
Let $V_{\Theta}$ and $V_{\$ }$ denote some given amounts of the
asset and the money, respectively. If at some time $t$ the assets
are exchanged in the proportion $V_{\$}:V_{\Theta}$ than we call
the number
$$p_{t}\equiv\ln \left( V_{\$} \right) - \ln \left( V_{\Theta} \right)
\leqno(6)$$ {\bf the logarithmic quotation} for the asset $\Theta
$. If the trader buys some amount of the asset $\Theta $ at the
quotation $p_{t_{1}}$ at the moment $t_{1}$ and sells it at the
quotation $p_{t_{2}}$ at the later moment $t_{2}$ then his profit
(or more precise the logarithmic rate of return) will be equal to
$$ r_{t_{1},t_{2}}=p_{t_{2}}-p_{t_{1}}. \leqno(7)$$
The logarithmic rate of return, contrary to $p_{t}$, does not
depend on the choice of unit used to measure the assets in
question. From the projective geometry point of view
$r_{t_{1},t_{2}}$ is an invariant and $p_{t}$ is not, cf the
discussion of demand and supply curves in the Section 4.

The {\bf merchandising mathematician model} (MM model) consists in
what we call the rational purchase followed by a random selling of
some asset $\Theta $. The rational purchase is simply a purchase
bound by a fixed  {\bf withdrawal price} $-a$ that is such a
logarithmic quotation for the asset $\Theta $, $-a$, above which
the trader gives the buying up.  The quotation method does not
matter to the process of rational purchase. A random selling can
be identified with the situation when the withdrawal price is set
to $-\infty $ (the trader in question is always bidding against
the rest of traders).

Let us suppose now that the model describes a stationary process,
that is the probability density  $\eta \left( p \right)$ of the
random variable $p$ (the logarithmic quotation) does not depend on
time. Note that it is sufficient to know the logarithmic
quotations up to arbitrary constant because what matters is the
profit and profit is always a difference of quotations. This is
analogous with the classical physics where only differences of the
potential matter (cf Newton's gravity). Therefore we can suppose
that expectation value of the random variable $p$ is equal to
zero, $E\left( p \right) =0$. We shall also suppose that the
market is large enough not to be influenced by our trader. Let the
expression $[sentence]$ takes value 0 or 1 if the $sentence$ is
false or true, respectively (Iverson convention), Graham, Knuth,
and Patashnik (1994). The mean time of a random transaction
(buying or selling, it is a matter of convention) will be denoted
by $\theta $. The value of $\theta $ is fixed in our model due to
the stationarity assumption. Besides, to eliminate paradoxes (e.g.
infinite profits during  finite time spreads) $\theta $ should be
greater than zero. Let $x$ denote the probability that the
rational purchase would not occur:
$$x\equiv E_{\eta}\left( \left[ p>-a\right] \right) . \leqno(8)$$ The
expectation value of the rational purchase time of the asset
$\Theta $ is equal to $$ \theta\left( \left( 1-x\right) +2x\left(
1-x\right) +3x^{2} \left( 1-x\right) +4x^{3} \left( 1-x\right)
+\ldots \right) . \leqno(9) $$ The ratio of the expected duration
of the whole buying-selling cycle and the expected time of a
random reverse transaction is given by
$$\begin{array}{rl}\frac{E_{\eta}\left( \tau \right)}{\theta} =& 1 + \left( 1-x\right) \sum
_{k=1}^{\infty} kx^{k-1}\\ =& 1 +  \left( 1-x\right) \frac{d}{dx}
\sum _{k=0}^{\infty} x^{k}\\ =& 1 +  \left( 1-x\right)
\frac{d}{dx} \frac{1}{1-x} = 1+ \frac{1}{1-x}\end{array} .
\leqno(10)$$ Therefore the  mean length of the whole cycle is
given by $$E_{\eta}\left( \tau \right) =\left( 1 + \left(
E_{\eta}\left( \left[ p\leq -a\right] \right) \right) ^{-1}
\right) \theta . \leqno(11)$$ The logarithmic rate of return for
the whole cycle is
$$ r_{t,t+\tau}=- p_{\rightarrow \Theta} +p_{\Theta \rightarrow} ,
\leqno(12)$$ where the random variable $p_{\rightarrow \Theta}$
(quotation at the moment of purchase) has the distribution
restricted to the interval $(- \infty ,-a]$:
$$\eta _{ \rightarrow \Theta} \left( p\right) = \frac{\left[ p\leq -a\right] }
{E_{\eta} \left( \left[ p\leq -a \right] \right) }\; \eta \left( p
\right) .\leqno(13)$$ The random variable $p_{\Theta \rightarrow}$
(quotation at the moment of selling) has the probability density
$\eta$, as the selling is at random. The expectation value of the
of the profit after the whole cycle is $$ \rho _{\eta}\left( a
\right)= \frac{-\int _{-\infty}^{-a}p\;\eta \left( p\right) dp}{1+
\int _{-\infty}^{-a}\eta \left( p\right) dp} , \leqno(14)$$ which
follows from  Equations (5) and (13). This function has very
interesting properties (we will often drop the subscript $\eta $
in the following text) stated as the Theorem 1.
\begin{theorem}
The maximal value of the function $\rho$ , $a_{max}$ , lies at a
fixed point of $\rho$, that is fulfills the condition $\rho \left(
a_{max}\right) = a_{max}$. Such a fixed point $a_{max}$ exists and
$a_{max} >0$.
\end{theorem}
{\bf Proof} \\
The fixed point condition:
$$ \frac{-\int_{-\infty}^{-a}p\;\eta\left( p\right) dp}{1+\int_{-\infty}^{-a}p\;\eta
\left( p\right) dp} =a \leqno(15)$$ can be rewritten as :
$$a \left( 1+\int_{-\infty}^{-a}p\;\eta
\left( p\right) dp\right) = -\int_{-\infty}^{-a} \left( p+a\right)
\eta \left( p\right) dp + \int_{-\infty}^{-a}a\; \eta \left(
p\right)dp .\leqno(16)$$ This leads to $$a=-\int _{-\infty}^{0}
p\;\eta \left( p-a \right) dp .\leqno(17)$$ The derivative of the
righthand side of the Equation. (17) is equal to $$-\int
_{-\infty}^{-a} \eta \left( p\right) dp \leqno(18) $$ and it is
obvious that the righthand side of the Equation (17) is a
non-increasing positive function of $a$ that tends to $0$ for $
a\rightarrow \infty$. Remember that we have supposed that the
expectation value of $p$ is equal to $0$ so that $\eta$ cannot
identically vanish for $p\leq 0$. To end the proof it sufficient
no notice that the vanishing of the derivative of the function
$\rho $ :
$$a\; \eta \left( a \right) \left( 1+\int_{-\infty}^{-a}p\;\eta
\left( p\right) dp\right) +  \eta \left( a \right)
\int_{-\infty}^{-a}  p \;\eta \left( p\right) dp = 0
 .\leqno(19)$$ is
exactly the fixed point condition (14) multiplied by $\eta \left(
a \right)$. So for a non-vanishing $\eta \left( a \right)$ the
proof ends here. If $\eta \left( a \right)$ vanishes then the
derivative of $\rho$: $$\frac{d\rho\left( a \right)}{da}
=\frac{a\;\eta \left( a \right) }{ 1+\int_{-\infty}^{-a} \eta
\left( p\right) dp} +\frac{\eta\left( a\right) \int_{-\infty}^{-a}
p\;\eta \left( p\right) dp}{\left(1+ \int_{-\infty}^{-a} \eta
\left( p\right) dp\right) ^{2}} \leqno(20)$$ is non-negative in
small vicinities of
$a$ and there is no extremum at $a$. \\

It might be useful to analyze an example here.
\begin{example}[normal distribution]
Let now $\eta \left( p, \sigma\right)$ be the standard normal
probability density with the variance $\sigma$ and expectation
value $\hat p$ of a random variable $p$ $$\eta \left( p, \sigma
\right) \equiv \frac{1}{\sqrt{2\pi}\sigma } \exp \left( -
\frac{\left( p- \hat p \right) ^{2} }{2\sigma ^{2}}\right)
\leqno(21) $$ In this case the expectation value of the profit
during a whole cycle $\rho \left( a, \sigma \right)_{normal} $ (we
have explicitly shown the dependence on the variance $\sigma $)
has a nice scaling property :
$$\rho
\left( a, \sigma \right) _{normal}= \sigma \rho \left( a, 1
\right)_{normal} , \leqno(22) $$ and it is sufficient to work out
the $\sigma=1$ case only. If this is the case we get the maximal
expectation value of the profit for $a=0.27603$. According to the
Theorem 1. the maximal expected profit is also equal to $0.27063$.
\end{example}

It is worth to notice that the condition (16) clearly shows what
the maximal possible profit is. \\

It tempting to claim that the function $ \rho $ is a contraction.
In a general case this is not true. Simple inspection reveals that
if the probability has a very narrow and high maximum then $\rho $
is not a contraction in the vicinity of the maximum. But for any
realistic probability density one can start at any value of $a$
and by iteration wind up in the fixed point (Banach fixed point
theorem). We skip the details because they are technical and
unimportant for the conclusions of the paper.

\section{ DEMAND AND SUPPLY CURVES}

The literature on economics including texts avoiding mathematical
formalism abounds in graphs and diagrams presenting various demand
and supply curves.  Blaug in Blaug (1985) quotes at least a
hundred of such diagrams. This illustrates the importance the
economists attach to them. Such approach is also possible in the
MM model. Let us consider the functions $$ F_{s} \equiv E_{\eta
_{1}} \left( \left[ \xi \leq x\right] \right) =\int _{-\infty}^{x}
\eta _{1}\left( p\right) dp ,\leqno(23)$$ and
$$ F_{d} \equiv E_{\eta _{2}} \left(
\left[ \xi \geq x\right] \right) =\int ^{\infty}_{x} \eta
_{2}\left( p\right) dp , \leqno(24)$$ where we have introduced
two, in general case different, appropriate probability density
$\eta _{1}$ and $\eta _{2}$. They may differ due to the  existence
of a monopoly, specific market regulations, taxes, cultural habits
and so on. Let us recall that one can find two ways of presenting
demand/supply curves in the literature. The first one (French) is
based on the assumption that the demand is a function of prices
and is usually referred to as the Cournot convention. The
Anglo-Saxon literature prefers the Marshall convention with
reversed roles of the coordinates. The demand (supply) is not
always a monotonic function of prices (cf. the discussed below
turning back of demand/supply curves) therefore the Marshall
convention seems to be less convenient (one cannot use the notion
of a function). The MM model with the price-like parameter $x$
refers to the Cournot convention. So, for a fixed value $x$ of the
logarithm of the price of an asset $\Theta $, the value of the
supply function $F_{s}(x)$ is given by the probability of the
purchase of a unit at the price $\leq e^{x}$. The asset would be
provided by everyone who is willing to sell it at the price
$e^{x}$ or lower. The function $F_{d}(x)$ can be interpreted in an
analogous way. If we neglect the sources of possible differences
between $\eta _{1}$ and $\eta _{2} $ and, in addition, suppose
that at any fixed price there are no indifferent traders (that is
everybody wants to sell or buy) then we can claim that $$ E_{\eta
_{1}}\left(  [ \zeta \leq x] \right) + E_{\eta _{2}}\left( [ \zeta
> x]\right) =1. \leqno(25)
$$ The differentiation of the Equation (25) leads to $\eta _{1}
=\eta _{2}$. Under these conditions the price $e^{y}$ for which
$E_{\eta _{1}}\left(  [ \zeta \leq y] \right) = E_{\eta
_{2}}\left( [ \zeta
>y ]\right) =\frac{1}{2} $ establishes the equilibrium price in the
classical meaning. This simply means that this price the most
frequent one. Recall that the MM model is scaled so that this
price is $e^{0}= 1$.

It would be instructive to analize  the problem from the
projective geometry point of view. In this approach the market is
described in the $N-$dimensional real projective space, $\Re
P^{N}$ that is $(N+1)-$dimensional vector space $\Re ^{N+1}$ (one
real coordinate for each asset) subjected to the equivalence
relation $v\sim \lambda v$ for $v\in \Re ^{N+1}$ and $\lambda \neq
0$. For example we identify all portfolios having assets in the
same proportions. The actual values can be obtained by rescaling
by $\lambda$. The details would be presented elsewhere. In this
context  separate profits gained by buying or selling are not
invariant (coordinate free). The profit $r_{t,t+\tau}$ gained
during the whole cycle is given by the logarithm of an appropriate
anharmonic (cross) ratio, Courant and Robbins (1996), and is an
invariant (e.g. its numerical value does not depend on units
chosen to measure the assets). The anharmonic ratio for four
points lying on a given line, $A, \ B,\ C,\ D$ is the double ratio
of lengths of segments $\frac{AC}{AB}:\frac{DC}{DB}$ and is
denoted by $[A,B,C,D]$. In our case the anharmonic ratio  in
question, $[\Theta , U_{\rightarrow \Theta}, U_{\Theta
\rightarrow} ,\$]$, concerns the pair of points:
$$ U_{\rightarrow \Theta}\equiv\left( \upsilon ,\upsilon \cdot
e^{p_{\rightarrow \Theta }},\ldots \right) \ and \ U_{\Theta
\rightarrow}\equiv\left( w ,w \cdot e^{p_{\Theta
\rightarrow}},\ldots \right)  \leqno(26)$$ and the pair $\Theta \
and \ \$ $. The last pair results from the crossing of the
hypersurfaces $\overline{\Theta}$ and $\overline{\$}$
corresponding to the portfolios consisting of only one asset
$\Theta $ or $\$ $, respectively and the line $U_{\rightarrow
\Theta}U_{\Theta \rightarrow}$. The dots represent other
coordinates (not necessary equal for both points). The line
connecting $U_{\rightarrow \Theta}$ and $ U_{\Theta \rightarrow}$
may be represented by the one-parameter family of vectors
$u(\lambda )$ with $\mu-$coordinates given by
$$u_{\mu}\left( \lambda\right) \equiv \lambda\left( U_{\rightarrow
\Theta} \right) _{\mu} +\left( 1-\lambda \right) \cdot \left(
U_{\Theta \rightarrow} \right) _{\mu} . \leqno(27) $$ This implies
that the values of $\lambda $ corresponding to the points $\Theta
$ and $\$ $ are given by the conditions: $$ u_{0}\left( \lambda
_{\$}  \right) =\lambda _{\$} \left( U _{\rightarrow \Theta}
\right) _{0} + \left( 1-\lambda _{\$} \right) \cdot \left( U _{
\Theta \rightarrow } \right) _{0} =0 \leqno(28) $$ and
$$u_{1}\left( \lambda _{\$}  \right) =\lambda _{ \Theta } \left( U
_{\rightarrow \Theta} \right) _{1} + \left( 1-\lambda _{ \Theta}
\right) \cdot \left( U _{ \Theta \rightarrow } \right) _{1} =0.
\leqno(29) $$ Substitution of the Equation (26) leads to $$
\lambda _{\$} =\frac{w}{w-\upsilon} \leqno(30) $$ and $$\lambda
_{\Theta}= \frac{we^{p_{\Theta \rightarrow}}}{we^{p_{\Theta
\rightarrow}} -\upsilon e^{p_{\rightarrow \Theta}} }.\leqno(31) $$
The calculation of the logarithm of the cross ratio $[\Theta ,
U_{\rightarrow \Theta}, U_{ \Theta \rightarrow}, \$]$ on the line
$U_{\rightarrow \Theta} U_{\Theta \rightarrow}$ leads to
$$\begin{array}{rl}\ln \left[\Theta , U_{\rightarrow \Theta},
U_{\Theta \rightarrow} ,\$ \right] = &\ln \left[
\frac{we^{p_{\Theta \rightarrow}}}{we^{p_{\Theta \rightarrow}}
-\upsilon
e^{p_{\rightarrow \Theta}} },1,0, \frac{w}{w-\upsilon} \right] \\
= & \ln \frac{v\;w\;e^{p_{\Theta \rightarrow}}}{v\;e^{p_{
\rightarrow \Theta}}\;w } =p_{\Theta \rightarrow} -
p_{\rightarrow\Theta}\end{array} \leqno(32)$$ which corresponds to
the formula (7).

Contrary to the classical economics the balance in the MM model
does not result in  uniform quotations (prices) for the asset
$\Theta $ but only in a stationarity of the supply and demand
functions $E_{\eta _{1}}\left( [\zeta \leq x] \right) $ and
$E_{\eta _{2}}\left( [\zeta \geq x] \right) $. Therefore the MM
model is not valid when the changes in the probabilities happens
during periods shorter or of the order of the mean time
transaction $\theta$. Of course the presented above  stochastic
interpretation of the supply and demand remains valid in such
situations. In addition we can consider piecewise decreasing
functions $F_{s}$ and  $F_{d}$. Such generalization requires that
these function cease to be probability distribution functions
because their derivatives (probability densities) are not positive
definite. This corresponds to  the effect of {\sl turning back} of
the supply and demand curves what happens for work supplies and
the Giffen goods, Stigler (1947). In the Marshall convention these
curves loose the function property. In the Cournot convention
these curves are diagrams of multivalued functions. In this way
negative probability densities (Wigner function) gain interesting
economics reason for the existence. Wigner functions emerged in
the quantum theory, Feynman (1987). By a choice of stochastic
process consistent with the MM model one can determine the
dynamics of such a model, cf Blaquiere (1980). Therefore we
suspect that the departure from the laws of supply and demand
might be the first known example of a macroscopic reality governed
by quantum-like rules. Such hypothetical quantum economics started
with the evidence given by Robert Giffen in the British
Parliament, Stigler (1947) would have earlier origin than the
quantum physics. It should be noted here that from the quantum
game theory point of view the Gauss distribution function is the
only supply (demand) curve that fulfills the physical
correspondence principle. The authors would devote a separate
paper to this very interesting problem.

Let us note that the distribution functions allow for correct
description of the famous Zeno paradoxes (when grains form a pile?
when you start to be bald?) because the introduction of
probabilities removes the original discontinuity. For example the
problem of morally right prices: if the price is low (state 0)
nobody wants to sell and if the price is high (state 1) everybody
wants to sell. Without the probability theory we are not able to
describe intermediate states which, in fact, are typical on the
markets. Does it suggest that the MM model can also be applied to
problems where there is a necessity of finding maximum (minimum)
of a profit intensity like parameter?

\section{CONCLUDING REMARKS}

We have discussed the model where the trader fixes the maximal
price he is willing to pay for the asset $\Theta $ and then after
some time sells it at random. One can easily reverse the buying
and selling strategies. If this is the case the formula (13)
should be modified to: $$ \rho \left( b\right)= \frac{\int
^{\infty}_{b}p\;\eta \left( p\right) dp}{1+ \int ^{\infty}_{b}\eta
\left( p\right) dp} , \leqno(33)$$ where $b$ denotes the minimal
acceptable price of $\Theta $ (that is below which the trader
gives up the selling). \\

It is interesting that the optimal behaviour of the trader consist
in fixing the withdrawal price below the mean quotation so that
the difference is exactly the profit expected during a mean
buying-selling cycle. If he or she manages to do so then the
optimal and stable position is reached and no further manoeuvring
is necessary. So the best strategy is the self-consistent
correction of the withdrawal price, cf Theorem 1. The existence of
such a mechanism is highly required in dynamical markets where the
distributions of quotations are continuously changing. Please note
that if we set finite withdrawal prices for both type of
transactions (buying and selling) the above simple recipe cease to
work. One might ask if this suggests that the two-way transactions
should be avoided? Or the only correct model is the one consisting
in random buying  followed by selling with fixed withdrawal price?
One might suggest (suspect?) that the later case is the only one
when it is possible to define the quotation distribution
relatively to the subsequent selling. This might be compared with
widely spread opinion among brokers that the moment of closing of
a open position is much more important than the actual moment of
opening this position (i.e. random buying). Of course the terms
selling and buying are conventional: while selling the asset
$\Theta$ one
buys $\$ $. \\

At the end we would like to note that the process of searching
optimal solutions and fixed points are the key issues of
contemporary mathematical economics, Debreu (1981). Such classical
results as generalised Brouwer theorem, Kakutani (1941),  and the
Brown-Robinson iteration, Robinson (1951),  are widely applied.
The proposed MM model combines both ideas. The authors envisage
the extension of the MM model to the randomized withdrawal price
cases which might also generalise the results of Piotrowski  and
S\l adkowski (2001), where thermodynamics of investors was
considered and the temperature of
portfolios was defined.\\

{\it Institute of Theoretical Physics, University of Bia\l ystok,\
Lipowa 41, Pl-15424 Bia\l ystok, Poland; e-mail:
ep@alpha.uwb.edu.pl,
\begin{center}
and
\end{center}

  Institute of Physics, University of Silesia,
Uniwersytecka 4, Pl-40007 Katowice, Poland, e-mail:
sladk@us.edu.pl}

\begin {center}
 REFERENCES
\end{center}

\newcounter{bban}

\begin{list}
{}{\usecounter{bban}\setlength{\rightmargin} {\leftmargin}}

\item  Billingsley, R. (1979): {\it Probability and Measure}. New York: J. Wiley
and Sons.
\item Blaquiere, A. (1980):  "Wave Mechanics as a Two-Player
Game", in {\it Dynamical Systems and Microphysics}. New York:
Springer-Verlag.
\item Blaug, M. (1985): {\it Economic Theory in retrospect}. Cambridge: Cambridge
 University Press.
\item Courant, R., and Robbins, H. (1996): {\it What is
Mathematics? An Elementary Approach to Ideas and Methods}. Oxford:
Oxford University Press.
\item  Debreu, G. (1981) in {\it Handbook of Mathematical
Economics}, vol. II, ed. by K. J. Arrow and M. D. Intriligator.
Amsterdam: Elsevier Science.
\item  Feynman, R. P. (1987): "Negative Probabilities in Quantum
Mechanics", in {\it Quantum Implications, Essays in Honour of D.
Bohm}, ed. by B. J. Hiley and F. D. Peat. London: Routledge \&
Kegan Paul.
\item  Graham, R. L., Knuth, D. E., and Patashnik, O. (1994): {\it Concrete
Mathematics}. Reading: Addison-Wesley.
\item Kakutani, S. (1941): "A Generalization of Brouwer's Fixed Point Theorem", {\it Duke
Mathematical Journal}, {\bf 8} 457-458.
\item  Piotrowski, E. W. (1999): "Profit Intensity - Model of a Rational Merchant",
{\it Statistical Review (Przegl\c ad Statystyczny)}, {\bf 46},
191-197.
\item  Piotrowski E. W., and S\l adkowski, J. (2001):
The Thermodynamics of Portfolios", {\it
Acta Physica Polonica}, {\bf B32} 597-608.
\item Resnick, S. I. (1998): {\it A Probability Path}. Boston: Birkhauser.
\item Robinson, J. (1951): "An Iterative Method of Solving a Game", {\it Annals
of Mathematics}, {\bf 54} 296-301.
\item Stigler, G. J. (1947): "Notes on the History of the Giffen Paradox", {\it
 Journal of Political Economy}, {\bf 55} 152-156 .
\end{list}


                                  \end{document}